\documentclass[aps,pra,twocolumn,showpacs,superscriptaddress,groupedaddress]{revtex4-1}
\usepackage[margin={2cm,2cm}]{geometry}
\usepackage{graphicx}
\usepackage{natbib}
\usepackage{amsthm}
\usepackage[english]{babel}

\usepackage{amsmath}
\usepackage{amsfonts}
\usepackage{amssymb}
\usepackage{subfig}
\usepackage{hyperref}
\usepackage[usenames, dvipsnames]{color}
\usepackage{bbold}
\usepackage{bbm}
\usepackage{amsfonts}
\usepackage{hyperref}
\usepackage[english]{babel}
\hypersetup{
    colorlinks=true,
    linkcolor=ForestGreen,
    filecolor=magenta,      
    urlcolor=cyan,
    citecolor=blue
}
\usepackage{epstopdf}
\usepackage{lipsum}

\begin{document}
 \title{A generalised framework for non-classicality of states II: \\
 Emergence of non locality and entanglement}
  \author{Soumik Adhikary}
  \email{soumikadhikary@physics.iitd.ac.in}
  \affiliation{Department of Physics, Indian Institute of Technology Delhi, New Delhi-110016, India.}

 \author{Sooryansh Asthana}
  \email{sooryansh.asthana@physics.iitd.ac.in}
  \affiliation{Department of Physics, Indian Institute of Technology Delhi, New Delhi-110016, India.}

  \author{V. Ravishankar}
  \email{vravi@physics.iitd.ac.in}
  \affiliation{Department of Physics, Indian Institute of Technology Delhi, New Delhi-110016, India.}

\date{\today}

\begin{abstract}
 A unified formalism was developed in \cite{Adhikary18}, for describing non classicality of states by introducing pseudo projection operators in which both quantum logic and quantum probability are naturally embedded. In this paper
 we show, as the first practical application, how non-locality and entanglement emerge as  two such important manifestations. It  provides  a perspective complementary to (i) the understanding of them that we have currently  \cite{BellInequalities, CHSHorig, Werner89}, and (ii) to  the algebraic approaches employed. The work also makes it  possible to obtain, in a systematic manner, an infinite number of conditions for non-classicality,
 for future applications.
 \end{abstract}

\maketitle


\section{\label{sec:level1}Introduction}
The idea of non-classicality of states is  a unique feature of quantum mechanics that distinguishes it 
from both classical physics and classical probability \cite{Dirac42, Bartlett45, Accardi81, Feynman87}.
Thus, it is  pivotal  to applications in quantum information and computation \cite{Benn84, Shor97,Benn93, Deutsch92}. For this reason, depending on the exigencies of situation, many standards of non classicality, such as  non-locality, entanglement, steering, and discord have been proposed \cite{BellInequalities, CHSHorig, Peres96, Wiseman07, Olliv01}. The need of the day is a  common unifying framework from which these many facets can be understood in a  natural fashion. It would also allow setting up criteria of non-classicality in a systematic fashion.

We have recently set up one such framework, hereafter denoted by I \cite{Adhikary18},  in which we introduce operators which we designate as pseudo projections. Non-classicality of states is captured by the associated pseudo probabilities  which admit negative values. The formalism,
which we recapitulate briefly in section \ref{prelim},  is entirely free of ad-hoc constructions,   and it has been shown  that  quantum logic
and quantum probability are inherent to the formalism. These were illustrated mainly with the example of two level systems.

As a direct continuation of I,  this paper shows how two important standards of non-classicality  -- non locality and entanglement, emerge as direct manifestations of quantum probability and quantum logic.
We show, with a judicious choice of classical logical propositions and/or   combinations of pseudo probabilities in a  given scheme, the emergence of Bell-CHSH non-locality in any dimension, and quantitative signatures for entanglement in
$2 \times 2$ dimensional spaces. Thereby,   this new demonstration also brings out yet another essential aspect of these important criteria.

\section{Preliminaries}
\label{prelim}

We begin with a quick recapitulation of pseudo projections and pseudo probabilities which were introduced in I.  

\subsection{Indicator function and their quantum representatives}
We start with  classical  observables $A, B,  \cdots $  defined  over a phase space $\Phi$ \footnote{More generally, it could be any sample space}.  Let $S^A_i \subset \Phi$ be the support for the outcome $A = a_i$. Similarly, let $S^B_j \subset \Phi$ be the support for the outcome $B = b_j$. If a system is in a state $f$, the respective probabilities for the outcomes will be given 
by the overlap of $f$ with the corresponding indicator functions $I_{S^A_i}$ and $I_{S^B_j}$. The indicator functions are boolean observables, taking  the value +1 within the support and zero  outside. For a given observable, the supports $S_i$ are mutually disjoint, and  partition  $\Phi$:  $S_i \bigcap S_j = \emptyset;~~ \bigcup\limits_i^{}S_i = \Phi$. 

Consider the transition to the quantum domain. The phase space maps to a Hilbert space $H$, and a classical state -- to a density operator $\rho$. An observable $A$  maps  to  a self adjoint operator denoted, again, by $A$ and which admits the spectral decomposition,  $A= \sum_ia_i\pi^A_i$;   the  eigen projections $\pi^A_i$ partition $H$ into a disjoint union of subspaces, ${\cal H} = \bigcup\limits_{i}^{}\oplus{\cal H}^A_i$. Finally,  the probability for the outcome $A=a_i$ is given by the overlap
$p^A_i = Tr(\rho \pi^A_i)$. All other outcomes are disallowed.

Two important mappings that follow are of particular relevance: 
$ I_{S^A_i} \rightarrow \pi_i;~~ S^A_i \rightarrow H_i$.  However, not all of these projections are the eigen projections of the operator. In fact, very many indicator functions would map to the trivial projection
$\pi=0$ (unless the spectrum is continuous), in which case, the corresponding supports $S_i \rightarrow \emptyset$, the null space.  

In short, the indicator functions either map to eigen projections, or their sums thereof, or to the trivial null operator. Similarly the supports map to either the eigen subspaces  or their direct sums thereof, or to the null set
$\emptyset$. There is no other possibility.

\subsection{Pseudoprojections for two observables}
The situation changes when joint outcomes of observables is considered. 
Classically, the probability for a joint outcome $A=a_i, B=b_j$ is equally easy to determine. One evaluates
the overlap of the classical state $f$ with the indicator function defined over the intersection of the corresponding supports $S_i \cap S_j$ (we suppress  the observable index henceforth, unless necessary).

The quintessential feature of quantum probability is that  $I_{S_i\bigcap S_j}$ does not map to a projection, unless the projections for the two outcomes commute. Assignment of  a joint probability is thus disallowed in most cases.
It does not forbid, however, the very question as to what the  quantum representative of $I_{S_i\bigcap S_j}$ is. Recall that, classically, $I_{S_i\bigcap S_j}$  is itself a Boolean observable. The answer to this question holds the key to formulate and understand the non-classical features of quantum probability.

 The answer may be inferred from rules of quantum mechanics.  The quantum representative is  just the symmetrised product of the projection operators:
  \begin{equation}
  I_{S_i\bigcap S_j} \equiv I_{S_i}I_{S_j} \rightarrow    {\bf \Pi}_{ij} = \frac{1}{2} (\pi_i\pi_j +\pi_j\pi_i).
  \end{equation}
   The quantum representative  ${\bf \Pi}_{ij}$ is hermitian, but not a projection,  unless $[\pi_i, \pi_j] =0$. This is but the simplest example of operators that we shall designate as pseudo projections. ${\bf \Pi}_{ij}$ possesses two essential properties: (i) As proved in I, it has at least one negative eigenvalue.   (ii) 
   The set, $ {\cal S}^2 =  \Big\{ {\bf \Pi}_{ij}\Big\}$,  of all pseudo projections corresponding to all possible joint outcomes of two observables, $A, B$  forms an over-complete set of operators and yields a resolution of identity,  as given by
\begin{eqnarray}
\label{Overcompleteness}
\sum_i\sum_j {{\bf \Pi}_{ij}}  &  =  & \sum_i\pi_i = \mathbb{1} \nonumber \\
\sum_j\sum_i{{\bf \Pi}_{ij}}  &  =  & \sum_j\pi_j = \mathbb{1} 
\end{eqnarray}
The proof is a  direct consequence of completeness of eigen projections of any observable.

\subsection{Pseudoprojection operators for multiple observables}
\label{pp2}
More generally, quantum representatives of indicator functions representing  joint outcomes of $N$ observables will be called pseudo projections. We are interested in the stuation when they are mutually incpmpatible. When $N \ge 3$, thet are not unique. 
Consider, thus, a conjunction of $N$ events $\Big\{ A^k = a^k_{i_k};~~ k=1,2,\cdots,N\Big\}$, where the index $k$ labels the observable.  If we were to denote the product of the corresponding projection operators in some order -- collectively denoted by the  ordered set \{$\alpha$\} --- by
${\cal A}^N_{\{\alpha\}}$, then the hermitised  sum 
\begin{equation}
{\bf \Pi}^N_{\{\alpha\}} =  \frac{1}{2} ({\cal A}^N_{\{\alpha\}}+ {\cal A}^{N \dag}_{\{\alpha\}}) 
\end{equation}
serves as a valid quantum representative of the classical indicator function, i.e., it is a pseudo projection.  We call this a unit pseudo projection. 

There are $\frac{N!}{2}$ such unit pseudo-projections, depending on the order in which the projection operators are arranged. The representative pseudo projection can be chosen to   be any element in the convex span of these unit pseudo projections:
\begin{equation}
\label{nu}
\sum_{\{\alpha\}} \lambda_{\{\alpha\}} {\bf \Pi}^N_{\{\alpha\}} \rightarrow {\bf \Pi}^N (\{\lambda_{\{\alpha\}}\})
\end{equation}
i.e., 
\begin{equation}
0 \le \lambda_{\{\alpha\}} \le 1;~ \sum_{\{\alpha\}} \lambda_{\{\alpha\}} =1.
\end{equation} 
 In general, each point in the manifold yields an inequivalent quantum representative of the underlying indicator function. The manifold would collapse to a point if all the projections were to commute. 

The richness afforded by this non uniqueness,  for
exploring fully all the aspects of quantum probability,  deserves a separate study.  For the present purposes, in this paper,  we employ,  
as in I,  only the completely symmetric combination, and denote it, generically,  by ${\bf \Pi}^N$.

Pseudo projections with $N$ observables also satisfy the following over completeness relation:
  Let ${\cal S}^N_{\{\alpha\}}$ be the set of all pseudo projections corresponding to all possible outcomes of the $N$ observables at hand. Let $\{\alpha\} = \{\beta, j\}$ where
$j$ refers to the outcomes of the observable labelled  $A^k$ and $\beta$ collectively denote the rest. Then,
\begin{equation}
\label{Marginal}
\sum_j{\bf \Pi}^N_{\{\beta, j\}} = {\bf \Pi}^{N-1}_{\{\beta \}}.
\end{equation} 
as in ${\cal S}^2$, it is a direct consequence of the completeness of eigen projections of observables. It may also be shown that unit pseudo-projections also possess negative eigenvalues.

\section{Pseudo Probability and Non Classicality}
As pointed out in I, pseudo projections generate pseudo probabilities. Let a system be in a state $\rho$. We then define the pseudo probability associated with a pseudo projection to be
\begin{equation}
 \mathcal{P}^N  = Tr\Big\{ \rho {\bf \Pi}^N\Big\}
 \end{equation}
  Since ${\bf \Pi}^N$ can admit negative eigenvalues, the corresponding pseudo probability can also be negative. We shall designate the set of all pseudo probabilities generated by a set ${\cal S}^N$ -- a pseudo probability scheme. 
 By virtue of relations in Eqs.(\ref{Overcompleteness}) and (\ref{Marginal}), it follows that pseudo probabilities in any scheme add up to one. 
 As a corollary of Eq.(\ref{Marginal}), pseudo probability schemes for subsets  of observables are just the marginals of the
 parent scheme.

Pseudo-probability schemes allow us to define non-classicality in a very broad sense. A state $\rho$ is deemed to be non-classical with respect to a set of observables $\{A_i\}$, even if a single entry in the corresponding pseudo probability scheme is negative. Conversely,  a state would be called classical with respect to the same set of observables iff all the entries in the scheme are non-negative.

As a corollary, sums of pseudo probabilities can also assume value ouside [0, 1], and can serve as
signatures of non classicality.

\subsubsection{Other logical operations}
It is convenient to set up the quantum representatives of indicator functions representing disjunction and negation, directly from the appropriate  suitable pseudo probability schemes, i.e., by employing standard probability rules. For example, the operator that corresponds to the disjunction, $A = a_i$ OR $B = b_j$,  can be obtained from the joint pseudo probability scheme  for  observables $A, B$, as follows:

\begin{align} 
{\bf R}_{a_i b_j}^{A B} & = \sum_{l} {\bf \Pi}_{a_i b_l}^{A B} +\sum_{l} {\bf \Pi}_{a_l b_j}^{AB} -  {\bf \Pi}_{a_i b_j}^{AB} \nonumber \\ 
&=  {\pi}_{a_i}^{A} + {\pi}_{b_j}^{B} -  {\bf \Pi}_{a_i b_j}^{AB}
\end{align}
Similarly, the quantum representative of the negation of an event is obtained by subtracting its representative pseudo projection from identity. 

Just as in  conjunction (joint events), and even more so, the expression for OR given above is prescriptive, and  not completely inferred.  This is not a drawback since other prescriptions reflect, again, the inherent ambiguity in the construction of quantum analogs of classical entities. A further discussion of this richness is, at this stage,  an unnecessary digression.

As in many other fortunate  situations,  it is  not always necessary to have a full knowledge of the scheme to derive some of the important tests for non-classicality. It is most certainly true of Bell CHSH non-locaility, as we show below. And again, many good tests for entanglement can be devised with a much smaller set of pseudoprobabilities.

We  devote the rest of the paper to demonstrate just this:  how non-locality and entanglement can be understood as natural manifestations of quantum logic and quantum probability, both of which are captured by pseudo probabilities.
The propositions involve pseudoprojections and quantum representatives of other logical operations involving disjunction and negation . These results go way beyond the ones obtained in I for single systems. 

\section{Notations and Pictorial Representations} 
 First, we establish some notations for the sake of compactness.
All observables, considered henceforth, are dichotomic, with eigenvalues  $\pm 1$. Thus, the two outcomes are negations of each other.  
The proposition $\mathcal{L} (A = +1)$ is denoted by $\mathcal{L} (A)$, and its negation, $\mathcal{L} (A = -1)$, by
 $\mathcal{L} (\overline{A})$. Conjunctions are written as simple juxtapositions, by omitting the sign $\wedge$. 
  However, the OR operation (disjunction) will be explicitly denoted by the standard symbol $\vee$.   We agree to separate observables belonging to different subsystems by a semicolon. The following example illustrates the notations.

 \begin{equation}
 \label{lpn}
 \mathcal{L} \{ A_1 = +1\} \wedge \mathcal{L} \{B_1 = -1\} \wedge  \mathcal{L} \{B_2 = +1 \} \equiv
 \mathcal{L} \{A_1; \overline{B}_1 B_2 \}
  \end{equation}
 Here, the observables $A_1$ and $B_{1,2}$ belong to the first and the second subsystems respectively. 
The pseudoprobability corresponding to $A = +1$ is denoted by $\mathcal{P} (A)$, and the one for its negation, $A = -1$, by
 $\mathcal{P} (\overline{A})$.
For example, the  pseudoprobability corresponding to the proposition in Eq.( \ref{lpn}) will be denoted by 
 $ \mathcal{P} \{A_1; \overline{B}_1 B_2 \}$.
 
\begin{figure}[!ht]
  \centering
  \includegraphics[width=0.5\textwidth]{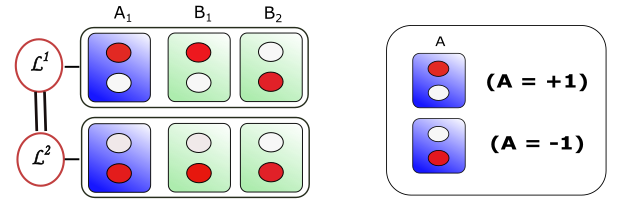}
  \caption{\label{Notation}Pictorial representation of the proposition $\mathcal{L}^1 \{A_1; B_1 \overline{B}_2 \} \vee \mathcal{L}^2 \{\overline{A}_1; \overline{B}_1 \overline{B}_2 \}$. Inset: The pictorial representation of the outputs $\pm 1$ of an observable. }
\end{figure}

 We also  introduce a pictorial representation alongside the algebraic  expressions for the logical propositions since the latter  can be lengthy, and not easily tractable. This is illustrated in Fig.(\ref{Notation}) where, the observables in the first and second subsystems are indicated by blue and green respectively. The outputs $\pm 1$ for an observable are indicated by a red button 
 in the appropriate slot. Entries in a given row refer to conjunction (AND) and different rows are related by disjunction (OR), as indicated by the double bond. 
 
\section{Non-locality and Entanglement from logical propositions}
 This section, together with the next,  contains  the main results of this paper.
 Our  approach here  is two pronged. In the first,  we demonstrate the break down of  elementary logical propositions  --- more precisely, of the associated probability, for  non locality  and entanglement. In the former case, we show the emergence of the Bell-CHSH inequality for all $ M \times N$ level systems. In the latter case, a set of witnesses are derived for $2 \times 2$ level systems.

  A word or two on how the propositions are constructed. We employ two guide lines. The propositions probe only the correlation space of the state by masking the nonclassicality coming from the subsystems. We further ensure that at least two or more noncommuting bases are involved which is essential to bring out the difference between separable and entangled states.

%

%
%

\subsection{Non-locality}
\begin{figure}[!ht]
  \centering
  \includegraphics[width=0.5\textwidth]{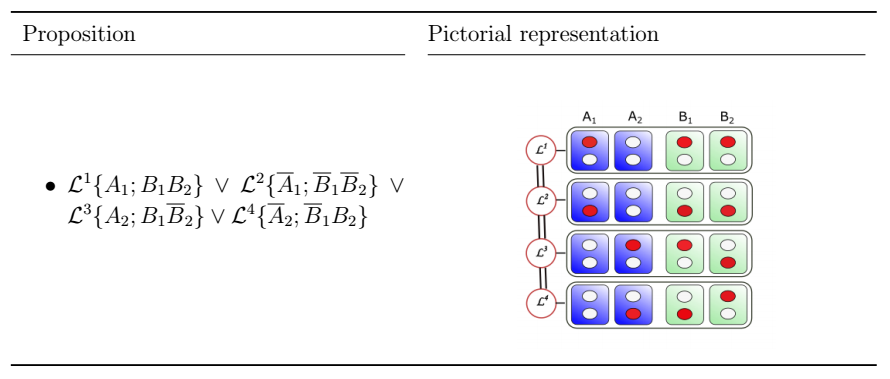}
  \caption{\label{12p} Figure showing the proposition underlying Bell-CHSH non-locality and its pictorial representation.}
\end{figure}

 Consider an $M \times N$ level system, and   pairs of  dichotomic observables,  $A_{1,2}$ and $B_{1,2}$,  belonging to respective subsystems.  The underlying  scheme $\mathcal{P}(A_1 A_2; B_1 B_2)$ consists of 16 entries. Of interest is  the simplest of the nontrivial propositions involving incompatible observables, which are shown  in Fig. (\ref{12p}) --- both algebraically and pictorially.

The supports $S_i$  for the classical outcomes for their corresponding  conjunctions $\mathcal{L}^i$ in Fig.(\ref{12p}) are mutually disjoint.  Accordingly,   its indicator function $I_S$  for the proposition in Fig.(\ref{12p}) 
 is  given by the sum of the corresponding indicator functions, 
\begin{equation}
{I}_S = \sum_{i=1}^4 {I}_{S_i}.
\end{equation}
The  quantum representative of $ I_S$ is the corresponding sum, ${\bf{\Pi}}_{N} =\sum_i{\bf \Pi}_i$ of the pseudo projections representing each conjunction.  Note that none of the pseudo-projections in the summand is a projection. A state $\rho$ would be classical with respect to this proposition if the corresponding pseudo-probability respects the bounds
\begin{equation}
0 \le Tr\Big[\rho {\bf \Pi}_{N}\Big] \le 1
\end{equation}
The identity, $\pi^A_{\pm 1} = \frac{1}{2}( 1 \pm A)$, immediately leads to the classic Bell-CHSH inequality

\begin{equation}
\label{bell}
\Big| \Big<A_1B_1+ A_1B_2+A_2B_1-A_2B_2 \Big>_{\rho}  \Big|  \leq 2 
\end{equation}

 \noindent {\it Comparison with other derivations}: This derivation essentially  identifies the inadmissibility  of standard Boolean rules to  operations on a set of   logical propositions. Thereby, it throws further light on the violation of the corresponding  rule of classical probability in the form expressed in  \cite{CHSHorig}.  Truly, violation of Bell - CHSH  inequality is autonomous of the kinematics of inertial frames.  Hence, it stands in stark contrast with the very first derivation which employed space like separations. This observation does,  by no means,    diminish  the deep physical and philosophical consequences that follow from combining non-locality  with special relativity. The derivation  also shows that 
    jointly measurable observables cannot lead to the violation of the Bell-CHSH inequality \cite{Fine82, Fine82-2,Khalfin85, Wolf09}. It will be seen  that this last conclusion continues to hold true for entanglement also.
   
   For future comparison, we recast  Eq. (\ref{bell}) for the two-qubit case in terms of correlations, for the special geometry
   $Tr(A_1 A_2) = Tr(B_1 B_2) = 0$. We employ the forms $A_{1,2} = \vec{\sigma}\cdot \hat{a}_{1,2}$ and $B_{1,2} = \vec{\Sigma}\cdot \hat{b}_{1,2}$ in terms of the Pauli bases in the respective subspaces.  Writing the normalised sum of vectors as $\hat{b} = \frac{1}{\sqrt{2}}(\hat{b}_1 + \hat{b}_2)$ and $\hat{b}' = \frac{1}{\sqrt{2}}(\hat{b}_1 - \hat{b}_2)$, we obtain the following inequality for nonlocality,
   \begin{equation}
\label{n1}
 \Big\vert \Big< \vec{\sigma} \cdot \hat{a}_1 \vec{\Sigma} \cdot \hat{b} + \vec{\sigma} \cdot \hat{a}_2 \vec{\Sigma} \cdot \hat{b}'\Big> \Big \vert_{\rho} >  \sqrt{2}.
\end{equation}

\subsection{Entanglement}
\label{ent}
 Though all non-local states are entangled,  the converse statement is not necessarily true\cite{Werner89}, suggesting that entanglement admits further refinements. Further, 
settling whether a state is entangled or not is  considered to be a hard problem,  which has  led to several  more modest approaches such as  majorisation relations, conditions based on correlation tensors and studies involving concurrence \cite{Nielsen01, Mintert04, Vicente11}. Rather than hunt for a single proposition that would
deliver a witness which is capable of detecting all entangled states,   we take up  two-qubit systems,  and systematically construct  two inequivalent logical propositions of increasing complexity and three combinations of appropriate pseudo-probabilities, and show that each of them yields an entanglement witness, emphasising the  breakdown of the validity of an underlying classical proposition.

\subsubsection{\bf Proposition 1}
Let $A $ stand for the observable $\vec{\sigma}\cdot \hat{a}$. Henceforth, we denote the observables in the first and the second subsystems  respectively by Latin and Greek symbols. The respective  Pauli operators will be denoted by $\sigma_i$ and $\Sigma_i$. The first proposition has the same number of observables as in non locality,
but with additional constraints. This leads to inclusion of more states in the set of entangled states.

\begin{figure}[!ht]
  \centering
  \includegraphics[width=0.5\textwidth]{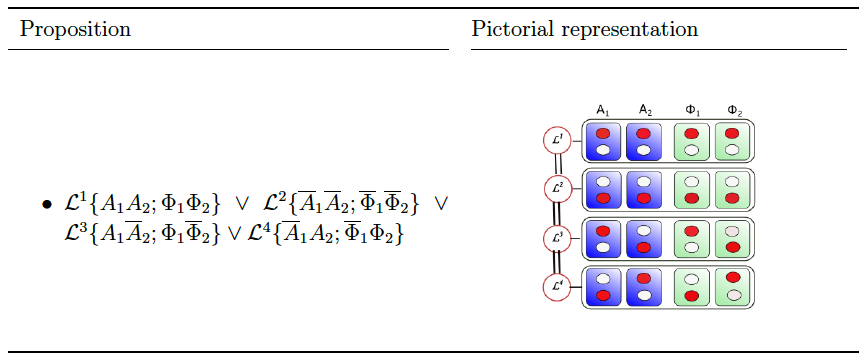}
  \caption{\label{22p} Proposition 1 and its  pictorial representation}
\end{figure}

Thus, let  two doublets of observables, $\{A_1, A_2\} $ and  $\{\Phi_1, \Phi_2\}$ belong to  the  first and  the second qubit respectively. They further obey    the orthonormality conditions $\hat{a}_1\cdot\hat{a}_2 = \hat{\phi}_1\cdot\hat{\phi}_2 =0$. The proposition,  which is more complex than the one for non-locality,   is displayed in Fig.(\ref{22p}), together with the attendant pictorial representation.

As with non-locality and all other subsequent examples, a state is classical if the associated pseudo probability is non-negative.   Mimicking  the steps just employed, we arrive at the first sufficiency condition for entanglement, which is  given by
\begin{equation}
\label{twotwo}
 \Big< \vec{\sigma} \cdot \hat{a}_1 \vec{\Sigma} \cdot \hat{\phi}_1 + \vec{\sigma} \cdot \hat{a}_2 \vec{\Sigma} \cdot \hat{\phi}_2 \Big>_{\rho} < -1.
\end{equation}
The correlations in the LHS of Eq. (\ref{twotwo}) are the same as in Eq.(\ref{n1}). But the   bound in the RHS renders  more states non-classical,  restating the fact that there are entangled states which are local. More importantly, it exhibits the logical distinction between local and non-local entangled states, complementing the view point emphasised in  \cite{Werner89}.
This distinction raises the possibility that inclusion of more observables (classical joint observations) may lead to even better witnesses and further logical distinctions within the family of entangled states.

 \subsubsection{\bf Proposition 2}
\begin{widetext}
  \begin{figure}[!ht]
  \centering
  \includegraphics[width=\textwidth]{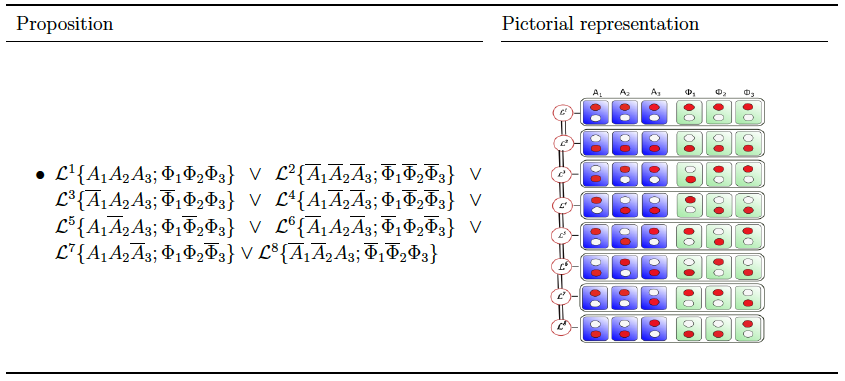}
  \caption{\label{33p} Proposition 2 and its  pictorial representation.}
\end{figure}
\end{widetext}

 We now consider two triplets each of three  orthonormal observables --  $\{A_i\},~\{\Phi_i\};~i \in \{1,2,3\}$ for the first  and second qubit respectively. More explicitly, $Tr(A_iA_j) = Tr(\Phi_i\Phi_j) = \delta_{i,j}$. The parent pseudo-probability scheme, $\mathcal{P}(A_1A_2A_3; \Phi_1\Phi_2\Phi_3)$,  would consist  of $2^6$ entries. But for our purposes,  it suffices to examine  a smaller combination of disjunctions shown in Fig.(\ref{33p}),involving only eight pseudo probabilities.
  
Yet again, for the same reasons stated above,  the pseudo projection for the unions  is simply the sum of each  pseudo projection for individual conjunctions. Violation  of the non negativity requirement for the associated pseudo probability for classicality  yields  the following sufficiency condition  for a state to be entangled:
\begin{equation} 
\label{ineq24Prop}
  \Big< \vec{\sigma} \cdot \hat{a}_1 \vec{\Sigma} \cdot \hat{\phi}_1 + \vec{\sigma} \cdot \hat{a}_2 \vec{\Sigma} \cdot \hat{\phi}_2 + \vec{\sigma} \cdot \hat{a}_3 \vec{\Sigma} \cdot \hat{\phi}_3 \Big>_{\rho}  <  -1
\end{equation}

The inequality (\ref{ineq24Prop}) is not new, and  has been derived earlier by G\"uhne et al. \cite{Gunhe03} and Werner \cite{ Werner89}. We note that their derivation  is driven by purely algebraic considerations, and that  their results were stated for one  specific geometry, 
$
\hat{a}_1 = \hat{\phi}_1 = \hat{x}; \ \  \hat{a}_2 = \hat{\phi}_2 = \hat{y}; \ \  \hat{a}_3 = \hat{\phi}_3 = \hat{z}
$,
 in contrast to our
approach which is motivated by violations of classical rules of probability associated with operations.  An even more direct derivation of this inequality will be given in the next section.

\section{Entanglement  from direct  violations of classical probability rules} 
\label{Sum_Composite} 
 We do away  with the task of explicit  formulation  in terms of underlying logical propositions and instead, deal with the entries in the pseudo probability schemes directly. 
  \subsubsection{Inequality 1}
 First, we
consider a pair of mutually orthogonal observables $A_1$, $A_2$ for the first qubit and
two triplets of  orthonormal observables $\{\Phi_1, \Phi_2, \Phi_3\}, \{\Theta_1, \Theta_2, \Theta_3\}$ for the second. To specify the detector geometry completely,  the normalised sums of the vectors, in the respective triplets $\{ \Phi_i \}$ and $\{ \Theta_i \}$, given by  $\hat{\phi} = \sum_{i=1}^3 \hat{\phi}_i/\sqrt{3}$ and $\hat{\theta} = \sum_{i=1}^3 \hat{\theta}_i/\sqrt{3}$, are chosen to be orthogonal:  $\hat{\phi}\cdot\hat{\theta} =0$. The number of pseudo probabilities in the scheme is $2^{8}$ and holds a wealth of information.  But for our purposes,  we may look at the sum of the  marginals of merely four  pseudo probabilities

\begin{eqnarray}
 \mathcal{S}_1 & = &  \mathcal{P}\{A_1;\Phi_1\Phi_2\Phi_3\} + \mathcal{P}\{\overline{A}_1;\overline{\Phi}_1\overline{\Phi}_2\overline{\Phi}_3\}  \nonumber \\
&+  &\mathcal{P}\{A_2;\Theta_1\Theta_2\Theta_3\} +
 \mathcal{P}\{\overline{A}_2;\overline{\Theta}_1\overline{\Theta}_2\overline{\Theta}_3\} .
\label{qp1}
\end{eqnarray}
 Classicality would force $\mathcal{S}_1$ to be non-negative.  Its violation yields a  sufficiency condition for entanglement given by
\begin{equation}
\label{e1}
 \Big< \vec{\sigma} \cdot \hat{a}_1 \vec{\Sigma} \cdot \hat{\phi} + \vec{\sigma} \cdot \hat{a}_2 \vec{\Sigma} \cdot \hat{\theta} \Big>_{\rho}  < -\frac{2}{\sqrt{3}},
\end{equation}

\subsubsection{Inequality 2}
The next  inequality  may be derived by enlarging the scheme further. For the  first subsystem,  we consider three orthonormal sets of   doublets  of observables $\{A_i\}$, $\{B_i\}$ 
 $\{C_i\}$.  For the second system,
  we choose three orthonormal sets of triplets of observables
 $\{\Phi_i\}, \{\Theta_i\}, \{\Psi_i\}$.  As in the earlier cases, the detector geometry is specified by requiring
 that both the  sets of normalised sums \Big\{$\hat{a}, \hat{b}, \hat{c}$\Big\}  and \Big\{$\hat{\phi}, \hat{\theta}, \hat{\psi}$\Big\}  
be  orthonormal sets. The sum of marginals of interest is
\begin{eqnarray}
\label{23}
 \mathcal{S}_2 & = &\mathcal{P}\{A_1 A_2;\Phi_1\Phi_2\Phi_3\} +\mathcal{P}\{\overline{A}_1\overline{A}_2;\overline{\Phi}_1\overline{\Phi}_2\overline{\Phi}_3\}\nonumber\\
& +&\mathcal{P}\{B_1 B_2;\Theta_1\Theta_2\Theta_3\} +\mathcal{P}\{\overline{B}_1\overline{B}_2;\overline{\Theta}_1\overline{\Theta}_2\overline{\Theta}_3\}\nonumber\\
&+&\mathcal{P}\{C_1 C_2;\Psi_1\Psi_2\Psi_3\} +\mathcal{P}\{\overline{C}_1\overline{C}_2;\overline{\Psi}_1\overline{\Psi}_2\overline{\Psi}_3\}
\end{eqnarray}
which leads to an independent sufficiency condition, i.e.,  a new  witness for non-separability,  which has the form
\begin{equation} 
\label{ineq23}
 \Big< \vec{\sigma} \cdot \hat{a} \vec{\Sigma} \cdot \hat{\phi} + \vec{\sigma} \cdot \hat{b} \vec{\Sigma} \cdot \hat{\theta} + \vec{\sigma} \cdot \hat{c} \vec{\Sigma} \cdot \hat{\psi} \Big><
-\sqrt{\frac{3}{2}}
\end{equation}

 \subsubsection{Inequality 3}
Finally, we consider three orthonormal triplets 
$\{A_i\}, \{B_i\}, \{C_i\}$ for the first qubit, and similarly  three orthonormal triplets of  $\{\Phi_i\}, \{\Theta_i\}, \{\Psi_i\}$ for the second.
As before, the normalised sums are chosen to be mutually orthogonal for each qubit. 
The convex sum of the marginals of pseudo probabilities, which is more involved, is  displayed in Eq. (\ref{33}).


\begin{eqnarray}
\label{33}
\mathcal{S}_3 &=&\mathcal{P}\{A_1 A_2 A_3;\Phi_1\Phi_2\Phi_3\} +\mathcal{P}\{\overline{A}_1\overline{A}_2 \overline{A}_3;\overline{\Phi}_1\overline{\Phi}_2\overline{\Phi}_3\}\nonumber\\
&+&\mathcal{P}\{B_1 B_2 B_3;\Theta_1\Theta_2\Theta_3\} +\mathcal{P}\{\overline{B}_1\overline{B}_2 \overline{B}_3;\overline{\Theta}_1\overline{\Theta}_2\overline{\Theta}_3\}\nonumber\\
&+&\mathcal{P}\{C_1 C_2 C_3;\Psi_1\Psi_2\Psi_3\} +\mathcal{P}\{\overline{C}_1\overline{C}_2 \overline{C}_3;\overline{\Psi}_1\overline{\Psi}_2\overline{\Psi}_3\}\nonumber\\
\end{eqnarray}

The resulting inequality yields an improved sufficiency condition for entanglement, given by 
$$ 
  \Big< \vec{\sigma} \cdot \hat{a} \vec{\Sigma} \cdot \hat{\phi} + \vec{\sigma} \cdot \hat{b} \vec{\Sigma} \cdot \hat{\theta} + \vec{\sigma} \cdot \hat{c} \vec{\Sigma} \cdot \hat{\psi} \Big>  <  -1 \nonumber
$$
This inequality is the same given in Eq. (\ref{ineq24Prop}), thus demonstrating that
 employing a scheme or looking for violations of propositions are but two equivalent approaches. It is entirely a matter of convenience as to which approach is to be used.

More pertinently,   this analysis shows that none of these non-classicality conditions would follow if all the entries in the underlying pseudo-probability scheme were non-negative. A seemingly similar approach involving quasi probabilities \cite{Puri12} does not yield witnesses since a prior knowledge of the state is assumed to determine if the state is entangled. 

Just as with the distinction between non-locality and entanglement, the inequalities derived in this section induce a further refinement in characterising entanglement, depending on the set of classical rules  violated by the states. We explore their interrelationship in greater detail in the next section. 

\section{Examples and discussion } 
\begin{figure}[!ht]
  \centering
  \includegraphics[width=0.5\textwidth]{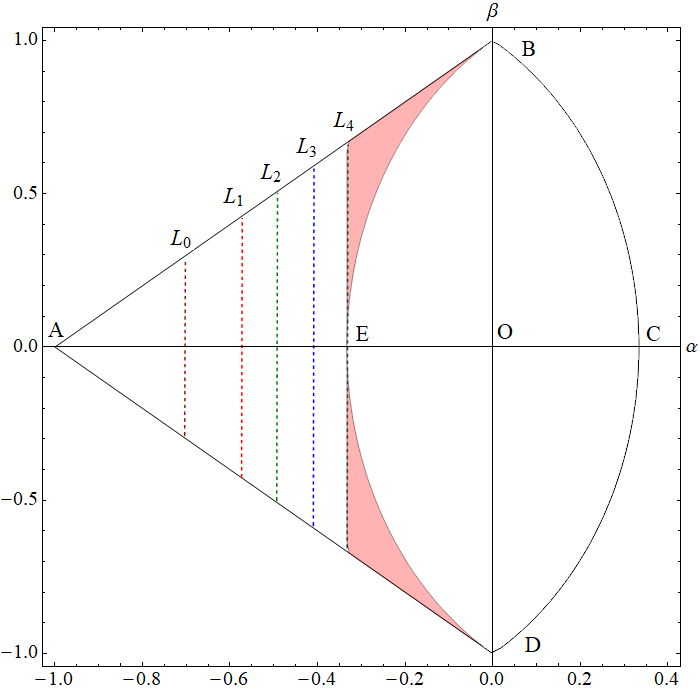}
  \caption{\label{illustration} Figure showing the relative strengths of the witnesses in the parameter space of $\rho$. 
  See text for details.}  
\end{figure}

We label the five  inequalities for entanglement,  \ref{n1}, \ref{e1}, \ref{twotwo}, \ref{ineq23} and \ref{ineq24Prop} as ${\bf W_0}, {\bf W_1}, {\bf W_2}, {\bf W_3}$ and ${\bf W_4}$ respectively. For illustration, and further discussion, we shall begin with
the special class of states, obtained by the addition of a local term to the Werner states,
\begin{equation}
\rho = \frac{1}{4} (1 + \alpha \vec{\sigma} \cdot \vec{\Sigma} + \beta \sigma_z).
\end{equation}
The main results are captured in Fig. (\ref{illustration}). The region bounded by the points  $A, B, C$ and $D$,  represents the space of all allowed states. The line $AC$ represents the Werner states, and the region contained between the arcs $BED$ and $BCD$ corresponds to separable states. The vertex $A$ represents the fully entangled singlet state, and the point $O$, the completely mixed state.

The five vertical lines mark the boundaries of   sets of entangled states (triangles with $A$ as their common vertex) detected by respective propositions. Of them, the first line $L_0$ marks the boundary between non-local and local states. The subsequent lines
represent, in order, the sets encompassed by the set of four sufficiency conditions, ${\bf W_i}$,  in the same order. The last line, $L_4$
encompasses the largest region, which includes all the entangled Werner states.  But it is still not exhaustive since it fails to detect entangled states in the region
shaded in pink. 
The existence of a logical proposition that would lead to a witness that detects all the entangled state is yet to be demonstrated.

\begin{figure}[!ht]
  \centering
  \includegraphics[width=0.4\textwidth]{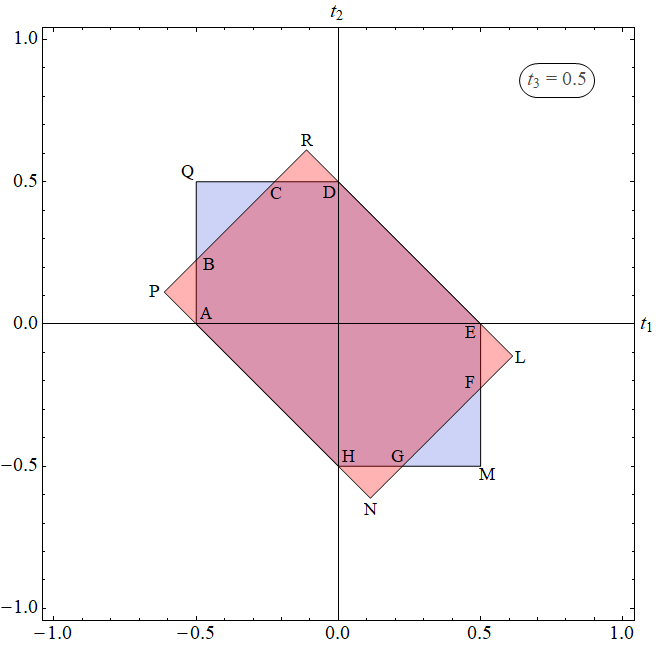}
  \caption{\label{strength} Figure showing a projection of the correlation space. See text for details.
     }
\end{figure}

It is possible to draw several strong conclusions for more general states. Consider an arbitrary two-qubit state in the SVD basis,
\begin{equation}
\rho = \frac{1}{4} (1 + \vec{P} \cdot \vec{\sigma} + \vec{Q} \cdot \vec{\Sigma} + \sum_{i=1}^3 t_{i} \sigma_i \Sigma_i).
\end{equation}
The witnesses, comprising entirely of correlations, are sensitive only to the singular
values $t_i$. Each witness, therefore detects entangled states in regions,  determined by the
corresponding set of inequalities imposed on the singular values. 
The region of the correlation space that represents a state is a tetrahedron, defined by the four inequalities \cite{Horodecki96},
 \begin{eqnarray}
 \label{posititvity}
  1 -  t_1 - t_2 - t_3 & \geq & 0  \nonumber \\
  1 - t_1 + t_2 +  t_3 & \geq & 0 \nonumber \\
  1 +  t_1 - t_2 + t_3 & \geq & 0 \nonumber \\
  1 +  t_1 + t_2 - t_3 & \geq & 0.
  \end{eqnarray}
First consider the witness ${\bf W_4}$. It partitions the correlation space into two parts. States that are separable and those that are entangled but evade detection by ${\bf W_4}$, satisfy the conditions:
\begin{eqnarray}
\label{W_4}
 1 +  t_1 + t_2 + t_3 & \geq & 0 \nonumber \\
  1 + t_1 - t_2 -  t_3 & \geq & 0 \nonumber \\
  1 -  t_1 + t_2 - t_3 & \geq & 0 \nonumber \\
  1 - t_1 - t_2 + t_3 & \geq & 0.
\end{eqnarray}
Together with conditions in Eq. (\ref{posititvity}), they  constitute the interior (and surface) of an octahedron.
 
The complementary region lying outside the octahedron corresponds to the entangled states detected by ${\bf W_4}$. This sufficiency condition also becomes necessary for the Bell diagonal states, as may be seen by employing partial transpose criterion, and as also displayed in Fig. (\ref{illustration}) for the special case of Werner states.

Thus, for the Bell diagonal states ${\bf W_4}$ is the strongest. But it still leaves the relative strengths of the four witnesses undetermined. To settle that we shall consider the other three witnesses. Starting with ${\bf W_3}$ we arrive at a new set of conditions:

\begin{eqnarray}
\label{W_3}
 \sqrt{\frac{3}{2}} +  t_1 + t_2 + t_3 & \geq & 0 \nonumber \\
  \sqrt{\frac{3}{2}} + t_1 - t_2 -  t_3 & \geq & 0 \nonumber \\
  \sqrt{\frac{3}{2}} -  t_1 + t_2 - t_3 & \geq & 0 \nonumber \\
  \sqrt{\frac{3}{2}} - t_1 - t_2 + t_3 & \geq & 0 
\end{eqnarray} 
which, together with Eq. (\ref{posititvity}) define a larger octahedron, which contains the separable and (undetected) entangled states. As before, $(t_1, t_2, t_3)$ corresponding to the entangled states detected by ${\bf W_3}$ must lie outside the octahedron. Between ${\bf W_4}$ and ${\bf W_3}$, the former is, of course, stronger. Of real interest, however, is to compare them with the conditions obtained by ${\bf W_{1,2}}$. These yield two sets of 12 bounds (that define dodecahedrons),

\begin{eqnarray}
\label{cond2}
c \pm t_1 \pm t_2 \geq 0 \nonumber \\
c \pm t_2 \pm t_3 \geq 0 \nonumber \\
c \pm t_3 \pm t_1 \geq 0
\end{eqnarray}
with $c = \frac{2}{\sqrt{3}} (1)$ for  ${\bf W_1 (W_2)}$. These 12 conditions, in conjunction with Eq. (\ref{posititvity}), yield the required region in the correlation space, that contains all the separable states and also some (undetected) entangled states. The states lying outside the respective regions are  all entangled and get detected.

With the  regions thus identified, we may immediately conclude that ${\bf W_4}$ is the strongest and that ${\bf W_3}$ is stronger than ${\bf W_1}$. We already know that ${\bf W_2}$ is stronger than ${\bf W_1}$. However, as may be seen in Fig. (\ref{strength}), ${\bf W_2}$ and ${\bf W_3}$ are mutually independent, since the entangled states that are detected have only partial overlaps. In  Fig. (\ref{strength}) we show projection of the correlation space, with $t_3 = 0.5$. The hexagon $QDEMHA$ and  the rectangle $PRLN$  constitute the set of  separable and undetected entangled states vis-a-vis ${\bf W_2}$ and ${\bf W_3}$, respectively. The overlap is the octagon $ABCDEFGH$. The triangles $QBC$ and $MGF$ are detected only by ${\bf W_3}$. Likewise the triangles $CDR, EFL, GHN, ABP$ are detected only by ${\bf W_2}$. These results, together with the examples discussed, reinforce the statement that different  witnesses reflect substructures in the space of entangled states that arise from violations of different classical rules.

\section{Conclusion}

In conclusion we show that, two important forms of non-classical correlations, {\it viz.}, non-locality and entanglement, can be naturally expressed through breakdown of classically valid logical propositions and hence violations of probability rules. Our construction automatically incorporates, the findings in \cite{Fine82, Khal92}, that jointly measurable observables cannot reveal non-locality of  a state, and naturally extends it to entanglement as well. 

The methods employed here do not exhaust ways of probing non-classicality. We have merely looked at convex sums of pseudo probabilities involving only correlations.  A fuller study of non classicality  would require  non-linear combinations of pseudo probabilities that do not ignore local terms,  especially for entanglement. But the present work does demonstrate that the framework developed in I provides a broad basis for exploring varieties of non classicality and their interrelationships.

\section*{Acknowledgement}

 Soumik and Sooryansh thank the Council for Scientific and Industrial Research (Grant no. - 09/086 (1203)/2014-EMR-I and 09/086 (1278)/2017-EMR-I) for funding their research.

\end{document}